\documentclass[12pt]{iopart}

\usepackage{iopams}

\usepackage{graphicx} 
\usepackage{dcolumn} 
\usepackage{bm} 
\usepackage{textcomp}

\usepackage{color}
\usepackage{braket}

\begin{document}

\title{Quantum-walk measurements with sequential imaging of the same atomic sample}

\author{Ryuta Yamamoto$^{1}$ and Takeshi Fukuhara$^{1,2}$}

\address{$^1$RIKEN Center for Quantum Computing~(RQC), Wako, Saitama 351-0198, Japan}
\address{$^2$Department of Applied Physics, Waseda University, Okubo, Shinjuku-ku, Tokyo 169-8555, Japan}
\ead{ryuta.yamamoto@a.riken.jp}

\date{\today}

\begin{abstract}
We demonstrate sequential single-atom-resolved imaging of the same ultracold atomic sample and apply this technique to quantum-walk measurements in an optical lattice.
In quantum gas microscope experiments, fluorescence imaging provides microscopic information on atomic distributions but induces motional heating through photon scattering, making it difficult to reuse the same sample for subsequent quantum dynamics.
To overcome this limitation, we apply Raman sideband cooling with parameters optimized for post-imaging cooling, enabling sequential single-atom-resolved measurements of the same sample.
By optimizing the cooling parameters with machine learning, we achieve high fidelity over multiple imaging cycles.
To demonstrate the capability of this technique, we measure one-dimensional quantum walks of single atoms in an optical lattice.
By determining the initial atomic positions from fluorescence images, we extract quantum-walk transition probabilities without requiring deterministic preparation of the initial atomic positions.
The observed dynamics agree with numerical simulations of quantum walks of single atoms.
Our method provides a route toward reducing experimental overhead in quantum gas microscope studies requiring repeated acquisition of spatially resolved correlation data.

\end{abstract}

\noindent{\it Keywords}: cold atoms, optical lattice, laser cooling, quantum gas microscope, quantum dynamics, machine learning

\maketitle

\section{Introduction}

Quantum walks are quantum-mechanical analogues of classical random walks, in which interference between different propagation paths leads to ballistic spreading and probability distributions that differ fundamentally from those of classical diffusion.
Owing to these features, quantum walks provide a versatile framework for investigating quantum transport, quantum search algorithms, topological phenomena, and the dynamics of interacting quantum systems~\cite{Kempe2003,Venegas-Andraca2012,Qiang2024QuantumApplication}.
They have been realized in a variety of physical platforms, including photonic systems, trapped ions, superconducting circuits, and ultracold atoms~\cite{Qiang2024QuantumApplication}.
Ultracold atoms in optical lattices offer a particularly clean and controllable platform for realizing quantum walks in real space. 
Quantum walks of neutral atoms in state-dependent optical lattices have been theoretically proposed and experimentally demonstrated using site-resolved fluorescence imaging~\cite{Dur2002QuantumLattices,Karski2009QuantumAtoms}.
The development of quantum gas microscopy enabled the preparation and observation of individual atoms in strongly correlated lattice systems~\cite{Weitenberg2011Single-spinInsulator}.
In particular, strongly correlated quantum walks of interacting bosons were observed with single-site resolution, revealing two-particle correlations and interaction-induced quantum dynamics~\cite{PPhilipp_QW:2015}.
Related single-site-resolved studies also observed the coherent propagation of a deterministically created spin impurity in a one-dimensional optical lattice, providing a direct probe of mobile spin-excitation dynamics in a many-body background~\cite{TFukuhara_QD:2013,TFukuhara:2013b,TFukuhara:2015}.
More recently, optical tweezer programmability has enabled two-dimensional quantum walks in the Hubbard regime, demonstrating the potential of neutral-atom platforms for programmable quantum-walk experiments and spatial-search protocols~\cite{Young2022Tweezer-programmableLattice}.

In quantum gas microscope experiments, single-atom-resolved fluorescence imaging requires the scattering of many photons~\cite{Bakr2009ALattice,Sherson2010}.
Although such imaging provides direct access to microscopic atomic distributions, it also leads to motional heating and can disturb the atomic sample.
To suppress this heating and maintain site localization during imaging, various cooling techniques, including optical molasses, Raman sideband cooling, and electromagnetically induced transparency cooling, have been implemented~\cite{Schaefer:2020}.
These techniques enable high-fidelity single-atom-resolved imaging in optical lattices and have become essential tools for quantum gas microscopy.
In addition, nondestructive or minimally destructive imaging schemes have demonstrated the possibility of extracting information from ultracold atomic gases while suppressing heating and atom loss~\cite{Patil2014NondestructiveGas}.
Nevertheless, cooling atoms to low-lying motional states after single-atom-resolved imaging, thereby enabling their reuse for subsequent quantum dynamics measurements, remains an important challenge.

Here, we demonstrate sequential single-atom-resolved imaging of the same atomic sample and apply this technique to one-dimensional quantum-walk measurements in an optical lattice.
Our measurement protocol employs Raman sideband cooling~(RSC) in two distinct stages: imaging RSC, which compensates for recoil heating during fluorescence imaging, and post-imaging RSC, which reduces motional excitations after imaging to prepare atoms for subsequent quantum dynamics.
By optimizing the parameters of post-imaging RSC, we enable repeated measurements on the same atomic sample without preparing a new sample after each image.
Since each image determines the atomic positions, the measured positions serve as the initial conditions for the subsequent quantum walk, eliminating the requirement for deterministic preparation of atoms at specific lattice sites.
A similar approach using measured initial atomic positions to investigate subsequent wave packet expansion has been demonstrated with fermionic atoms in continuous space~\cite{Verstraten2025WP_ContinuousSpace}.
While previous studies have demonstrated single-atom-resolved measurements of quantum dynamics, our work emphasizes repeated measurements over multiple imaging cycles using the same atomic sample.
The observed quantum-walk dynamics are consistent with numerical simulations incorporating residual motional excitations, supporting the effectiveness of post-imaging cooling.
Our approach to repeated measurements using the same atomic sample reduces the need for repeated sample preparation and offers a route toward more efficient acquisition of spatially resolved correlation data in quantum gas microscope experiments.

\section{Sequential measurements with post-imaging cooling}
The experiments are performed using a triangular-lattice quantum gas microscope for $^{87}\mathrm{Rb}$ atoms, in which single-site-resolved fluorescence imaging is realized with RSC~\cite{Yamamoto2020Single-site-resolvedLattice}.
As described in the introduction, imaging RSC and post-imaging RSC serve different purposes and therefore require different cooling parameters. During imaging RSC, a sufficiently high photon-scattering rate is required for efficient fluorescence detection.
However, in the ideal limit of RSC, atoms accumulate in a dark state associated with the vibrational ground state, suppressing further photon scattering.
Thus, imaging RSC must balance efficient cooling against the requirement to collect sufficient fluorescence photons, rather than simply maximizing the vibrational ground-state population~\cite{Yamamoto2020Single-site-resolvedLattice}.
Consequently, atoms retained after fluorescence imaging are not necessarily prepared in sufficiently low motional states for subsequent coherent quantum dynamics.
We therefore optimize the parameters of post-imaging RSC separately to achieve more efficient motional cooling.

\begin{figure}[t]
    \centering
    \includegraphics[width=1\linewidth]{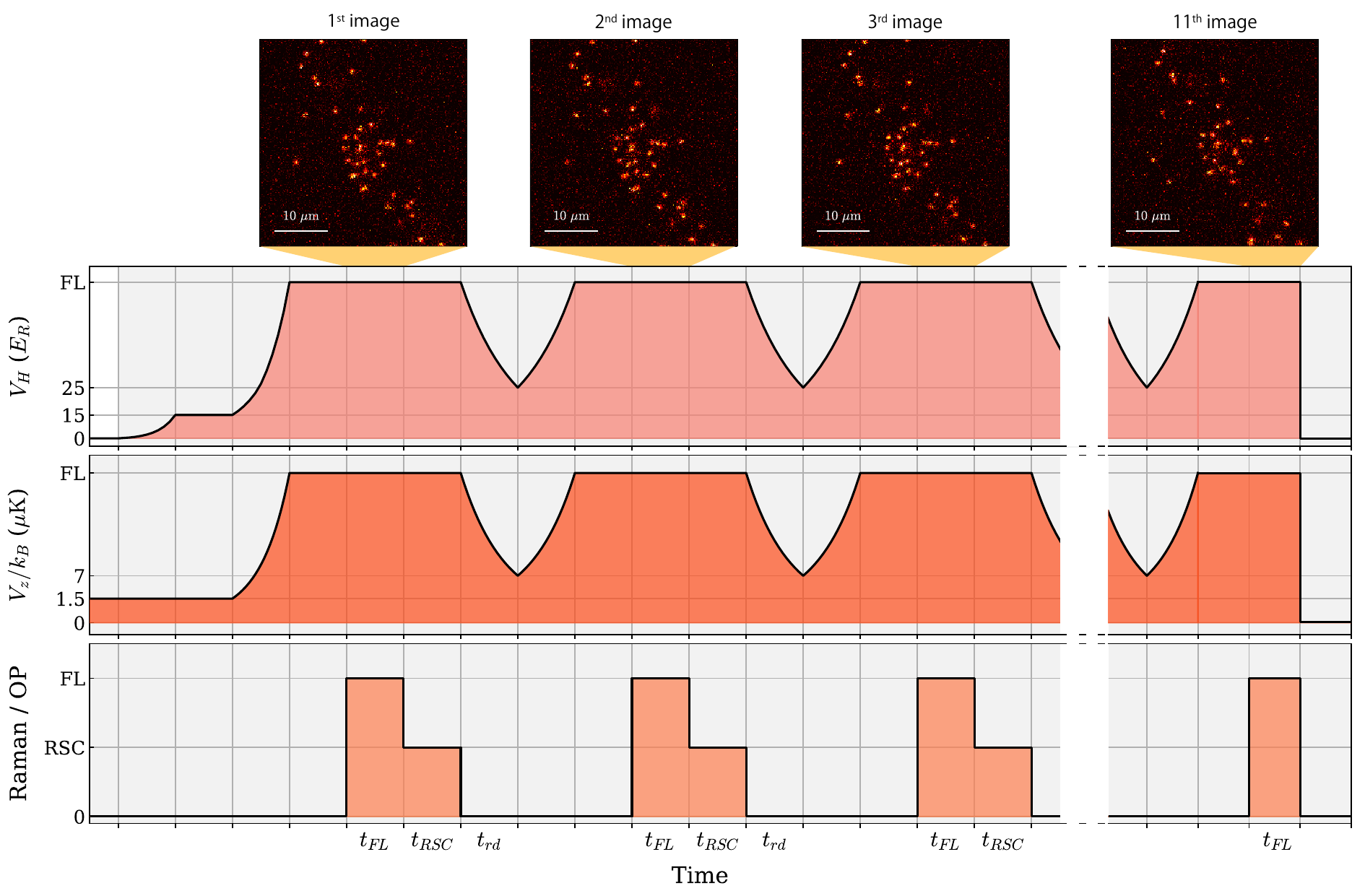}
    \caption{
        Schematic of the experimental sequence for repeated fluorescence imaging with post-imaging cooling.
        $V_{H}$ and $V_{z}$ denote the depths of the horizontal and vertical lattices, respectively, while Raman/OP indicates the intensities of the Raman and optical pumping beams.
        The recoil energy $E_R = \hbar^2 k_L^2/2m$ is defined for the wavelength $\lambda = 2 \pi/k_L =1064$~nm of the lattice beams. 
        The durations $t_\mathrm{FL}$, $t_\mathrm{RSC}$, and $t_\mathrm{rd}$ correspond to fluorescence imaging, post-imaging RSC, and lattice ramp-down to the shallow lattice potential, respectively.
        The respective durations are 500~ms, 150~ms, and 800~$\mu\mathrm{s}$.
        The labels ``FL" and ``RSC" on the vertical axes indicate the parameter values used for fluorescence imaging and post-imaging RSC, respectively. 
        Note that the vertical and horizontal scales are only for visualization and do not represent exact scaling.
    }
    \label{fig:sequential_images}
\end{figure}

During fluorescence imaging, the optical lattice is sufficiently deep to suppress atomic tunneling and maintain site localization, making it difficult to evaluate the effectiveness of post-imaging cooling from the observed atomic distributions.
To evaluate the performance of post-imaging cooling, we therefore lower the lattice depths after imaging and cooling, making atoms in higher motional states more likely to tunnel away from their initial sites, while those in low-lying motional states remain localized.
By repeating this procedure, we evaluate the ability of post-imaging cooling to maintain atomic localization over multiple imaging cycles.
A schematic of the experimental sequence is shown in figure~\ref{fig:sequential_images}.
Throughout this work, the number of fluorescence images is fixed at 11.
We optimized the RSC parameters using Bayesian optimization, following the approach employed in our previous work~\cite{INakamura_MLO:2019}.
The optimization score is defined as the ratio of the fluorescence counts in the central region of the $11^\mathrm{th}$ image to those in the corresponding region of the first image.
Maximizing this ratio favors cooling conditions that maintain atomic localization throughout repeated imaging cycles, thereby enabling the same atomic sample to be reused for subsequent quantum dynamics measurements without repeated sample preparation.
The optimization results are shown in \ref{Parameter_table_of_post-imaging_RSC}.

\begin{figure}[t]
    \centering
    \includegraphics[width=0.75\linewidth]{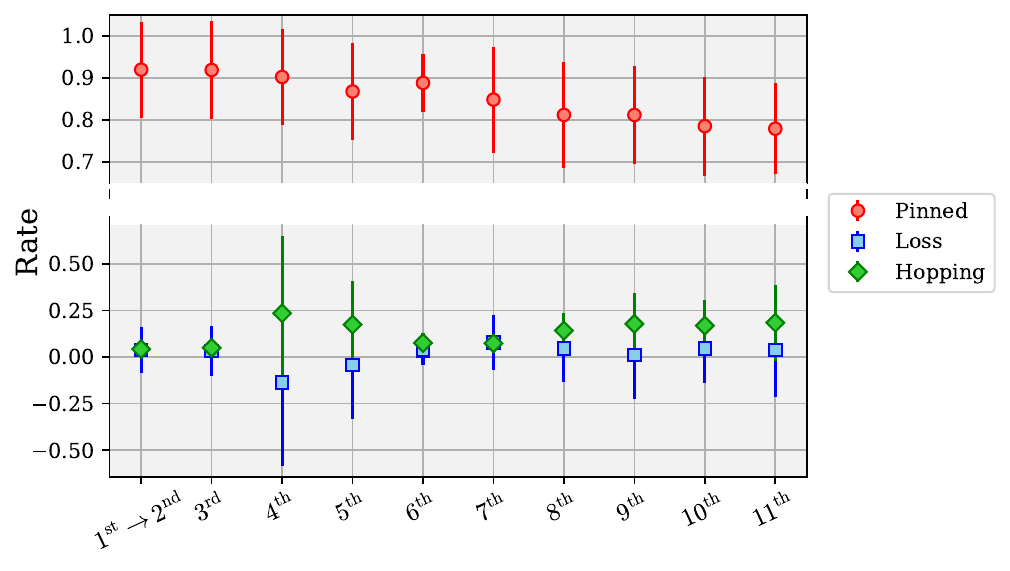}
    \caption{
        Performance of repeated single-atom-resolved imaging evaluated by comparing the first and $n^\mathrm{th}$ fluorescence images. 
        The horizontal triangular-lattice depths are ramped down to $V_{12}=V_{23}=V_{31}=25~E_R$, and the vertical lattice depth is reduced to $V_z/k_B = 7~\mathrm{\mu K}$.
        Here, $V_{ij}$ denotes the lattice depth arising from the interference between the horizontal lattice beams $i$ and $j$, which are two of three beams forming the triangular optical lattice. 
        Red circles, blue squares, and green diamonds represent the pinned, loss, and hopping rate, respectively.
        Error bars represent the standard deviation across independent experimental sequences.
    }
    \label{fig:Fidelity_sequenctial}
\end{figure}

To evaluate the performance of sequential imaging, we compare the atom positions detected in the first and $n^\mathrm{th}$ fluorescence images~(figure~\ref{fig:Fidelity_sequenctial}). 

We define the pinned rate as the fraction of atoms detected at a given lattice site in the first image that are detected again at the same site in the $n^\mathrm{th}$ image.
The loss rate is defined as the fractional decrease in the number of detected atoms from the first to the $n^\mathrm{th}$ image, while the hopping rate corresponds to the fraction detected at different lattice sites.
To account for the slight drift of the lattice-site positions between successive images, we correct the measured atomic positions as described in \ref{Lattice_Phase_shift}.

As shown in figure~\ref{fig:Fidelity_sequenctial}, the pinned rate remains high over repeated imaging cycles.
Even after the $11^\mathrm{th}$ image, the pinned rate is $0.78\pm0.11$, while the loss and hopping rates are $0.04\pm0.25$ and $0.18\pm0.20$, respectively.
The average pinned rate between successive images is $0.93\pm0.03$, indicating a high probability of retaining atoms at the same lattice sites between consecutive measurements.
Interestingly, the value expected from independently applying this average rate over ten cycles, $(0.93)^{10}\approx0.48$, is substantially lower than the measured pinned rate of $0.78$.
This discrepancy may arise from errors in determining atomic positions to lattice sites, including those associated with the correction for lattice-site drift~(see \ref{Lattice_Phase_shift}).
Such errors can produce apparent hopping events in successive image pairs, leading to an underestimation of the pinned rate between consecutive images.

These results demonstrate that post-imaging RSC enables repeated single-atom-resolved imaging of the same atomic sample while suppressing atom loss and hopping.

\section{Quantum-walk measurements with sequential imaging}

Having optimized the post-imaging cooling parameters, we demonstrate the applicability of sequential imaging to measurements of coherent quantum dynamics by observing one-dimensional quantum walks of single atoms in an optical lattice.
Specifically, we repeatedly perform quantum-walk evolution and single-atom-resolved imaging on the same atomic sample, using the measured atomic positions as the initial conditions for each subsequent quantum walk. 
The single-particle quantum walk is described by the Hamiltonian
\begin{equation}
    \hat{\mathcal{H}} = -J\sum_{i} \left( \hat{a}^\dagger_{i+1} \hat{a}_{i} + \hat{a}^\dagger_{i} \hat{a}_{i+1} \right)
    + \sum_i E_i \hat{a}^\dagger_i \hat{a}_i,
\end{equation}
where $\hat{a}^\dagger_i$ and $\hat{a}_i$ are the bosonic creation and annihilation operators at lattice site $i$, respectively, $J$ is the nearest-neighbor tunneling energy, and $E_i=m\omega^2 x_i^2/2$ represents the external harmonic confinement.
Here, $m$ is the atomic mass, $\omega$ is the trapping frequency, and $x_i=i a_{\mathrm{lat}} + x_\mathrm{offset}$ denotes the position of site $i$, with lattice spacing $a_{\mathrm{lat}}$ and offset $x_\mathrm{offset}$.

The wave function after a quantum-walk evolution time $t_\mathrm{QW}$ is given by
\begin{equation}
    \ket{\psi(t_\mathrm{QW})} =
    \exp\left[-i\frac{\hat{\mathcal{H}}}{\hbar}t_\mathrm{QW}\right] \ket{\psi(0)}, 
\end{equation}
where $\hbar=h/2\pi$, with $h$ being the Planck constant.

\begin{figure}[h]
    \centering
    \includegraphics[width=1\linewidth]{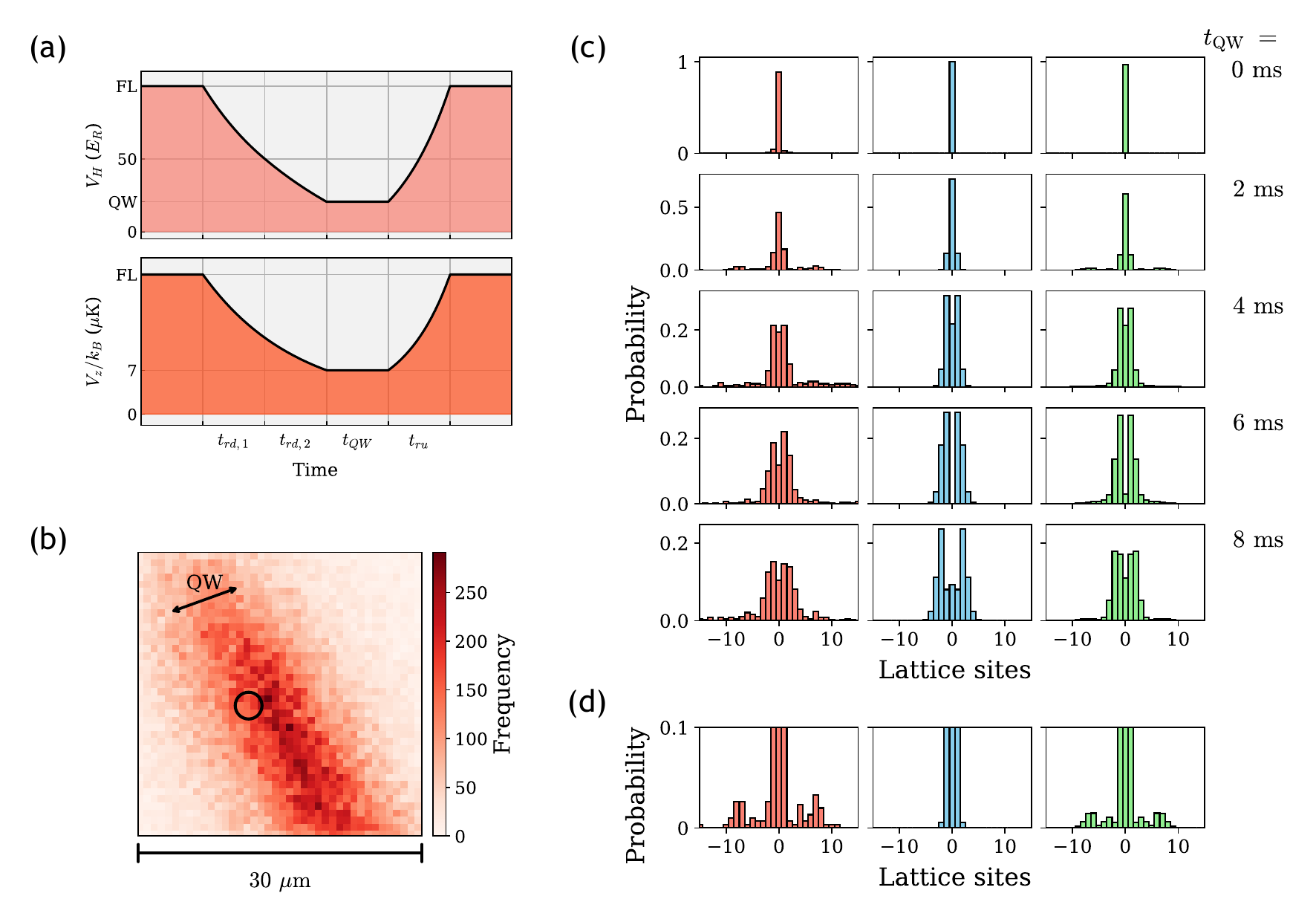}
    \caption{
        Single-atom quantum walks observed using sequential imaging.
        (a) Schematic of the experimental sequence for quantum-walk measurements.
        The durations $t_\mathrm{rd,1}$, $t_\mathrm{rd,2}$, and $t_\mathrm{ru}$ correspond to the first and second stages of the lattice ramp-down to the shallow lattice potential, and ramp-up to the deep lattice potential for imaging, respectively.
        These durations are 300~$\mu\mathrm{s}$, 500~$\mu\mathrm{s}$, and 200~$\mu\mathrm{s}$, respectively.
        The duration $t_\mathrm{QW}$ is varied to probe the quantum-walk dynamics at different evolution times.
        The labels "FL" and "QW" on the vertical axes indicate the lattice depths during fluorescence imaging and quantum walk dynamics, respectively. 
        (b) Initial atomic distribution.
        The black circle indicates the ROI near the trap center used for the analysis, and the double-headed arrow indicates the axis along which the quantum walk occurs.
        (c) Quantum-walk distributions for different evolution times $t_{\mathrm{QW}}$.
        The left column shows the experimental results, while the middle column shows numerical simulations under ideal quantum-walk conditions, assuming that all atoms occupy the motional ground state and neglecting the external harmonic potential.
        The right column shows numerical simulations incorporating both the motional ground and first excited states with populations of $83\%$ and $14\%$, respectively, as well as the external harmonic potential.
        The calculated distributions in the right column are averaged over the ROI shown in (b) with equal weighting.
        (d) Quantum-walk distributions at $t_{\mathrm{QW}}=2~\mathrm{ms}$ shown in (c), plotted on an expanded vertical scale to highlight the contribution from atoms in the first excited motional state.
    }
    \label{fig:QW_single}
\end{figure}

To realize effectively one-dimensional quantum walks in the triangular optical lattice, we set the horizontal lattice depths to $V_{12}=V_{23}=2.5~E_R$ and $V_{31}=25~E_R$ (where $E_R=\hbar^2 k_L^2/2m$ is the recoil energy for the lattice wavelength
$\lambda=2\pi/k_L=1064~\mathrm{nm}$), while maintaining the vertical lattice depth at $V_z/k_B=7~\mu\mathrm{K}$ (figure~\ref{fig:QW_single}(a)).
This lattice configuration creates an array of one-dimensional tubes. 
Under these conditions, tunneling along the vertical direction is negligible owing to the large lattice spacing and strong confinement.
The nearest-neighbor tunneling energies along the one-dimensional direction are $J_0/h=31.3$~Hz for atoms in the ground state and $J_1/h=-335~\mathrm{Hz}$ for atoms in the first excited state~(\ref{Calc_tunneling_energies}). 
For the analysis of quantum-walk dynamics, we select pairs of successive fluorescence images and include only those one-dimensional tubes in which exactly one atom is detected in both images.
The atomic position in the first image serves as the initial position for the quantum walk, while that in the second image determines the final position after the evolution.
Quantum-walk distributions were measured by varying the evolution time $t_{\mathrm{QW}}$. 

Due to the relatively strong harmonic confinement with the trapping frequency $\omega/2\pi\sim 79~\mathrm{Hz}$, the quantum-walk dynamics depend on the initial atomic position within the trap.
Consequently, simply averaging the data over all initial positions washes out the characteristic interference features of the quantum-walk distributions.
To mitigate this effect, we first concentrated the initial atomic distribution near the trap center by increasing the intensity of the additional optical trapping potential.
We then restricted the analysis to a region of interest~(ROI) around the trap center, as indicated in figure~\ref{fig:QW_single}(b).
Only one-dimensional tubes with an atom located within the ROI, as determined by the first of two successive images, were selected.
The resulting quantum-walk distributions are shown in figure~\ref{fig:QW_single}(c).

The measured distributions exhibit evolution similar to that predicted for atoms in the motional ground state, although noticeable discrepancies remain.
We attribute these differences mainly to two effects: residual motional excitations and the influence of harmonic confinement.

First, because post-imaging cooling is not perfectly efficient, a fraction of the atoms remains in excited motional states.
These atoms exhibit faster tunneling dynamics due to the larger magnitude of their tunneling energies, resulting in more rapid spatial spreading than for atoms in the motional ground state.
Raman sideband spectroscopy indicates a ground-state population of approximately 83\%~(see \ref{Parameter_table_of_post-imaging_RSC}).
To take this effect into account, we calculate the quantum-walk dynamics for atoms in the ground and first excited motional states and combine the resulting distributions with weights of 83\% and 14\%, respectively. 
The remaining 3\% population in higher motional states is neglected, and the resulting quantum-walk distribution is not renormalized, yielding a total probability of 97\%.

Second, although restricting the analysis to the ROI reduces the influence of harmonic confinement, the quantum-walk dynamics still depend on the initial atomic position.
Averaging over different initial positions therefore smooths out the interference features in the measured distributions.
We incorporate this effect into the numerical simulations by averaging the calculated quantum-walk distributions over the initial positions within the ROI.

By including both residual motional excitations and the position dependence arising from harmonic confinement, the numerical simulations reproduce the experimental distributions reasonably well, as shown in the right column of figure~\ref{fig:QW_single}(c).

\begin{figure}[t]
    \centering
    \includegraphics[width=1\linewidth]{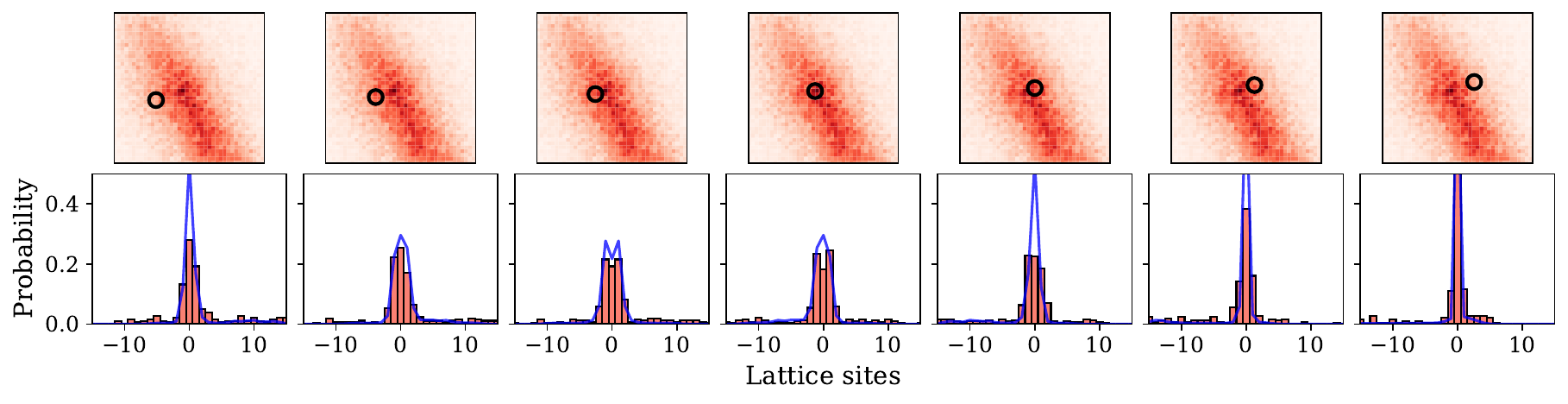}
    \caption{
        Position-dependent quantum-walk dynamics in the external harmonic trap.
        The top panels show the initial atomic distribution, with black circles indicating the post-selected initial positions.
        The bottom panels show the corresponding quantum-walk distributions after an evolution time of $t_{\mathrm{QW}}=4~\mathrm{ms}$~(histograms).
        The solid blue lines show numerical simulations incorporating residual motional excitations and the external harmonic potential, using the same model as in figure~\ref{fig:QW_single}(c).
        }
    \label{fig:Effect_ext_trap}
\end{figure}

We further investigate the influence of harmonic confinement on the quantum-walk dynamics by comparing the measured distributions after a fixed evolution time of $t_{\mathrm{QW}}=4~\mathrm{ms}$ for different initial atomic positions.
As shown in figure~\ref{fig:Effect_ext_trap}, the quantum-walk distributions strongly depend on the post-selected initial positions, revealing the effect of site-dependent energy offsets on the tunneling dynamics.
As the initial position moves farther from the trap center, the energy differences between adjacent lattice sites increase, thereby suppressing tunneling and reducing the spatial spreading of the atoms.

Repeated quantum walks followed by position measurements can also lead to a gradual broadening of the atomic distribution, resembling the diffusive spreading of a classical random walk.
Although the evolution during each quantum walk is coherent, the position measurement between successive walks removes the coherence between different lattice sites, resulting in stochastic atomic displacements over repeated cycles.
Consequently, the fraction of atoms remaining near the trap center may decrease with successive measurements, reducing the statistics available for quantum-walk analysis within the central ROI.
This represents a practical limitation of repeated quantum-walk measurements in the presence of harmonic confinement.

\section{Conclusion}
In conclusion, we have demonstrated repeated single-atom-resolved imaging of ultracold $^{87}\mathrm{Rb}$ atoms in an optical lattice using post-imaging Raman sideband cooling.
By optimizing the post-imaging cooling parameters, we achieved a high probability of retaining atoms at their original lattice sites over repeated imaging cycles, allowing the same atomic sample to be reused for subsequent measurements without preparing a new ensemble after each image.

To demonstrate the applicability of this technique to measurements of quantum dynamics, we performed one-dimensional quantum-walk experiments using sequential imaging.
Since the initial atomic positions are identified from the fluorescence images, these measurements do not require deterministic preparation of atoms at specific lattice sites.
The observed quantum-walk distributions exhibit characteristic coherent spreading and are reasonably reproduced by numerical simulations that account for residual motional excitations and the position dependence arising from harmonic confinement.
We also directly observed the dependence of quantum-walk dynamics on the initial atomic position, revealing the suppression of tunneling caused by site-dependent energy offsets in the external trapping potential.
These results demonstrate that post-imaging cooling enables repeated measurements of coherent quantum dynamics using the same atomic sample.

Our approach may be useful for investigating more complex tunneling dynamics in optical lattices.
One possible direction is to combine sequential imaging with Floquet engineering~\cite{Lignier:2007,Struck:2013,Ozawa2023ObservationLattices}.
In particular, periodic driving that breaks time-reversal symmetry can introduce complex tunneling amplitudes characterized by Peierls phases, providing access to effective magnetic fluxes and chiral dynamics~\cite{Barbiero2023FrustratedLadders}.
Single-atom quantum-walk measurements could offer a way to characterize such engineered tunneling processes and explore their connection to theoretical proposals for frustrated systems with synthetic gauge fields.

Another promising direction is measurement-based feedback control of quantum dynamics.
Recent theoretical studies have proposed quantum Maxwell's demons that exploit quantum fluctuations to generate directed particle transport, as well as feedback protocols exhibiting nontrivial topological properties~\cite{Liu:2026,Nakagawa:2025,Wen:2026}.
By combining repeated single-atom-resolved measurements with post-imaging cooling and measurement-dependent control of optical potentials, our system could provide an experimental platform for exploring these proposals.
In particular, the repeated quantum-walk measurements demonstrated here offer a natural starting point for investigating feedback-controlled quantum transport.

\ack
We thank Keegan McGehee for helpful discussions and carefully reading the manuscript. 
This work has been supported by JST ERATO (Grant Number JPMJER2302), JST Moonshot R\&D Program (Grant Numbers JPMJMS2269 and JPMJMS256D), JST ASPIRE (Grant Number JPMJAP24C2) and JSPS KAKENHI (Grant Numbers JP23K13049 and JP23K25830).
We acknowledge the stimulating discussions in the meeting of the Cooperative Research Project of the Research Institute of Electrical Communication, Tohoku University.

\section*{References}
\bibliographystyle{iopart-num}
\bibliography{references_QW} 

@article{Verstraten2025WP_ContinuousSpace,
    title = {{<i>In Situ</i> Imaging of a Single-Atom Wave Packet in Continuous Space}},
    year = {2025},
    journal = {Physical Review Letters},
    author = {Verstraten, Joris and Dai, Kunlun and Dixmerias, Maxime and Peaudecerf, Bruno and de Jongh, Tim and Yefsah, Tarik},
    number = {8},
    month = {2},
    pages = {083403},
    volume = {134},
    url = {https://link.aps.org/doi/10.1103/PhysRevLett.134.083403},
    doi = {10.1103/PhysRevLett.134.083403},
    issn = {0031-9007},
    arxivId = {2404.05699}
}

@article{Lignier:2007,
    title = {{Dynamical Control of Matter-Wave Tunneling in Periodic Potentials}},
    year = {2007},
    journal = {Physical Review Letters},
    author = {Lignier, H. and Sias, C. and Ciampini, D. and Singh, Y. and Zenesini, A. and Morsch, O. and Arimondo, E.},
    number = {22},
    month = {11},
    pages = {220403},
    volume = {99},
    url = {https://link.aps.org/doi/10.1103/PhysRevLett.99.220403},
    doi = {10.1103/PhysRevLett.99.220403},
    issn = {0031-9007}
}

@article{Struck:2013,
    title = {{Engineering Ising-XY spin-models in a triangular lattice using tunable artificial gauge fields}},
    year = {2013},
    journal = {Nature Physics},
    author = {Struck, J. and Weinberg, M. and {\"{O}}lschl{\"{a}}ger, C. and Windpassinger, P. and Simonet, J. and Sengstock, K. and H{\"{o}}ppner, R. and Hauke, P. and Eckardt, A. and Lewenstein, M. and Mathey, L.},
    number = {11},
    month = {11},
    pages = {738--743},
    volume = {9},
    publisher = {Nature Publishing Group},
    url = {https://www.nature.com/articles/nphys2750},
    doi = {10.1038/nphys2750},
    issn = {1745-2473}
}

@article{Liu:2026,
    title = {{Maxwell's demon for quantum transport}},
    year = {2026},
    journal = {Physical Review A},
    author = {Liu, Kangqiao and Nakagawa, Masaya and Ueda, Masahito},
    number = {2},
    month = {2},
    pages = {022436},
    volume = {113},
    url = {https://link.aps.org/doi/10.1103/rhvv-yd6w},
    doi = {10.1103/rhvv-yd6w},
    issn = {2469-9926}
}

@article{TFukuhara:2013b,
    title = {{Microscopic observation of magnon bound states and their dynamics}},
    year = {2013},
    journal = {Nature},
    author = {Fukuhara, Takeshi and Schau{\ss}, Peter and Endres, Manuel and Hild, Sebastian and Cheneau, Marc and Bloch, Immanuel and Gross, Christian},
    number = {7469},
    pages = {76--79},
    volume = {502},
    publisher = {Nature Publishing Group},
    url = {http://dx.doi.org/10.1038/nature12541},
    isbn = {1476-4687 (Electronic){\textbackslash}r0028-0836 (Linking)},
    doi = {10.1038/nature12541},
    issn = {00280836},
    pmid = {24067608},
    arxivId = {1305.6598}
}

@article{INakamura_MLO:2019,
    title = {{Non-standard trajectories found by machine learning for evaporative cooling of 87Rb atoms}},
    year = {2019},
    journal = {Optics Express},
    author = {Nakamura, Ippei and Kanemura, Atsunori and Nakaso, Takumi and Yamamoto, Ryuta and Fukuhara, Takeshi},
    number = {15},
    month = {7},
    pages = {20435},
    volume = {27},
    url = {https://doi.org/10.1364/OE.27.020435},
    doi = {10.1364/OE.27.020435},
    issn = {1094-4087}
}

@article{Ozawa2023ObservationLattices,
    title = {{Observation of chiral-mode domains in a frustrated XY model on optical triangular lattices}},
    year = {2023},
    journal = {Physical Review Research},
    author = {Ozawa, Hideki and Yamamoto, Ryuta and Fukuhara, Takeshi},
    number = {4},
    month = {11},
    pages = {L042026},
    volume = {5},
    publisher = {American Physical Society},
    url = {https://link.aps.org/doi/10.1103/PhysRevResearch.5.L042026},
    doi = {10.1103/PhysRevResearch.5.L042026},
    issn = {2643-1564},
    arxivId = {2304.07930}
}

@article{TFukuhara_QD:2013,
    title = {{Quantum dynamics of a mobile spin impurity}},
    year = {2013},
    journal = {Nature Physics},
    author = {Fukuhara, Takeshi and Kantian, Adrian and Endres, Manuel and Cheneau, Marc and Schau{\ss}, Peter and Hild, Sebastian and Bellem, David and Schollw{\"{o}}ck, Ulrich and Giamarchi, Thierry and Gross, Christian and Bloch, Immanuel and Kuhr, Stefan},
    number = {4},
    month = {2},
    pages = {235--241},
    volume = {9},
    url = {http://www.nature.com/doifinder/10.1038/nphys2561},
    doi = {10.1038/nphys2561},
    issn = {1745-2473}
}

@article{Kempe2003,
    title = {{Quantum random walks: An introductory overview}},
    year = {2003},
    journal = {Contemporary Physics},
    author = {Kempe, Julia},
    number = {4},
    pages = {307--327},
    volume = {44},
    doi = {10.1080/00107151031000110776},
    issn = {00107514},
    arxivId = {quant-ph/0303081}
}

@article{Venegas-Andraca2012,
    title = {{Quantum walks: a comprehensive review}},
    year = {2012},
    journal = {Quantum Information Processing},
    author = {Venegas-Andraca, Salvador Elías},
    number = {5},
    pages = {1015--1106},
    volume = {11},
    url = {http://www.springerlink.com/index/10.1007/s11128-012-0432-5},
    doi = {10.1007/s11128-012-0432-5},
    issn = {1570-0755}
}

@article{Sherson2010,
    title = {{Single-atom-resolved fluorescence imaging of an atomic Mott insulator}},
    year = {2010},
    journal = {Nature},
    author = {Sherson, Jacob F and Weitenberg, Christof and Endres, Manuel and Cheneau, Marc and Bloch, Immanuel and Kuhr, Stefan},
    number = {7311},
    month = {9},
    pages = {68--72},
    volume = {467},
    publisher = {Nature Publishing Group},
    url = {http://www.ncbi.nlm.nih.gov/pubmed/20720540 http://www.nature.com/doifinder/10.1038/nature09378},
    doi = {10.1038/nature09378},
    issn = {0028-0836},
    pmid = {20720540}
}

@article{TFukuhara:2015,
    title = {{Spatially Resolved Detection of a Spin-Entanglement Wave in a Bose-Hubbard Chain}},
    year = {2015},
    journal = {Physical Review Letters},
    author = {Fukuhara, Takeshi and Hild, Sebastian and Zeiher, Johannes and Schau{\ss}, Peter and Bloch, Immanuel and Endres, Manuel and Gross, Christian},
    number = {3},
    month = {7},
    pages = {035302},
    volume = {115},
    url = {https://doi.org/10.1103/PhysRevLett.115.035302},
    doi = {10.1103/PhysRevLett.115.035302}
}

@article{PPhilipp_QW:2015,
    title = {{Strongly correlated quantum walks in optical lattices}},
    year = {2015},
    journal = {Science},
    author = {Preiss, Philipp M. and Ma, Ruichao and Tai, M. Eric and Lukin, Alexander and Rispoli, Matthew and Zupancic, Philip and Lahini, Yoav and Islam, Rajibul and Greiner, Markus},
    number = {6227},
    month = {3},
    pages = {1229--1233},
    volume = {347},
    url = {https://doi.org/10.1126/science.1260364},
    doi = {10.1126/science.1260364},
    issn = {0036-8075},
    arxivId = {1409.3100v2}
}

@article{Wen:2026,
    title = {{Symmetry and Topology of Successive Quantum Feedback Control}},
    year = {2026},
    journal = {Physical Review Letters},
    author = {Wen, Junxuan and Gong, Zongping and Sagawa, Takahiro},
    number = {9},
    pages = {090802},
    volume = {136},
    publisher = {American Physical Society},
    url = {https://doi.org/10.1103/48k3-tb1j},
    doi = {10.1103/48k3-tb1j},
    issn = {10797114},
    pmid = {41861323}
}

@article{Schaefer:2020,
    title = {{Tools for quantum simulation with ultracold atoms in optical lattices}},
    year = {2020},
    journal = {Nature Reviews Physics},
    author = {Sch{\"{a}}fer, Florian and Fukuhara, Takeshi and Sugawa, Seiji and Takasu, Yosuke and Takahashi, Yoshiro},
    number = {8},
    pages = {411--425},
    volume = {2},
    doi = {10.1038/s42254-020-0195-3},
    issn = {25225820},
    arxivId = {2006.06120}
}

@article{Nakagawa:2025,
    title = {{Topology of Discrete Quantum Feedback Control}},
    year = {2025},
    journal = {Physical Review X},
    author = {Nakagawa, Masaya and Ueda, Masahito},
    number = {2},
    month = {4},
    pages = {021016},
    volume = {15},
    publisher = {American Physical Society},
    url = {https://link.aps.org/doi/10.1103/PhysRevX.15.021016},
    doi = {10.1103/PhysRevX.15.021016},
    issn = {2160-3308}
}

@article{Young2022Tweezer-programmableLattice,
    title = {{Tweezer-programmable 2D quantum walks in a Hubbard-regime lattice}},
    year = {2022},
    journal = {Science},
    author = {Young, Aaron W. and Eckner, William J. and Schine, Nathan and Childs, Andrew M. and Kaufman, Adam M.},
    number = {6608},
    pages = {885--889},
    volume = {377},
    doi = {10.1126/science.abo0608},
    issn = {10959203},
    pmid = {35981010},
    arxivId = {2202.01204}
}

@article{Bakr2009ALattice,
    title = {{A quantum gas microscope for detecting single atoms in a Hubbard-regime optical lattice}},
    year = {2009},
    journal = {Nature},
    author = {Bakr, Waseem S. and Gillen, Jonathon I. and Peng, Amy and F{\"{o}}lling, Simon and Greiner, Markus},
    number = {7269},
    month = {11},
    pages = {74--77},
    volume = {462},
    url = {http://www.nature.com/doifinder/10.1038/nature08482},
    isbn = {0028-0836},
    doi = {10.1038/nature08482},
    issn = {0028-0836},
    pmid = {19890326},
    arxivId = {0908.0174}
}

@article{Barbiero2023FrustratedLadders,
    title = {{Frustrated magnets without geometrical frustration in bosonic flux ladders}},
    year = {2023},
    journal = {Physical Review Research},
    author = {Barbiero, Luca and Cabedo, Josep and Lewenstein, Maciej and Tarruell, Leticia and Celi, Alessio},
    number = {4},
    month = {10},
    pages = {L042008},
    volume = {5},
    url = {https://link.aps.org/doi/10.1103/PhysRevResearch.5.L042008},
    doi = {10.1103/PhysRevResearch.5.L042008},
    issn = {2643-1564}
}

@article{Patil2014NondestructiveGas,
    title = {{Nondestructive imaging of an ultracold lattice gas}},
    year = {2014},
    journal = {Physical Review A},
    author = {Patil, Y. S. and Chakram, S. and Aycock, L. M. and Vengalattore, M.},
    number = {3},
    month = {9},
    pages = {033422},
    volume = {90},
    url = {http://link.aps.org/doi/10.1103/PhysRevA.90.033422},
    doi = {10.1103/PhysRevA.90.033422},
    issn = {1050-2947},
    arxivId = {arXiv:1404.5583v1}
}

@article{Qiang2024QuantumApplication,
    title = {{Quantum Walk Computing: Theory, Implementation, and Application}},
    year = {2024},
    journal = {Intelligent Computing},
    author = {Qiang, Xiaogang and Ma, Shixin and Song, Haijing},
    month = {1},
    pages = {0097},
    volume = {3},
    publisher = {American Association for the Advancement of Science},
    url = {https://spj.science.org/doi/10.34133/icomputing.0097},
    doi = {10.34133/icomputing.0097},
    issn = {2771-5892}
}

@article{Karski2009QuantumAtoms,
    title = {{Quantum Walk in Position Space with Single Optically Trapped Atoms}},
    year = {2009},
    journal = {Science},
    author = {Karski, Michal and F{\"{o}}rster, Leonid and Choi, Jai-Min and Steffen, Andreas and Alt, Wolfgang and Meschede, Dieter and Widera, Artur},
    number = {5937},
    month = {7},
    pages = {174--177},
    volume = {325},
    url = {https://www.science.org/doi/10.1126/science.1174436},
    doi = {10.1126/science.1174436},
    issn = {0036-8075}
}

@article{Dur2002QuantumLattices,
    title = {{Quantum walks in optical lattices}},
    year = {2002},
    journal = {Physical Review A},
    author = {D{\"{u}}r, W. and Raussendorf, R. and Kendon, V. M. and Briegel, H.-J.},
    number = {5},
    month = {11},
    pages = {052319},
    volume = {66},
    url = {https://link.aps.org/doi/10.1103/PhysRevA.66.052319},
    doi = {10.1103/PhysRevA.66.052319},
    issn = {1050-2947}
}

@article{Yamamoto2020Single-site-resolvedLattice,
    title = {{Single-site-resolved imaging of ultracold atoms in a triangular optical lattice}},
    year = {2020},
    journal = {New Journal of Physics},
    author = {Yamamoto, Ryuta and Ozawa, Hideki and Nak, David Can and Nakamura, Ippei and Fukuhara, Takeshi},
    number = {12},
    month = {12},
    pages = {123028},
    volume = {22},
    publisher = {IOP Publishing},
    url = {https://iopscience.iop.org/article/10.1088/1367-2630/abcdc8},
    doi = {10.1088/1367-2630/abcdc8},
    issn = {1367-2630},
    arxivId = {2010.05200}
}

@article{Weitenberg2011Single-spinInsulator,
    title = {{Single-spin addressing in an atomic Mott insulator}},
    year = {2011},
    journal = {Nature},
    author = {Weitenberg, Christof and Endres, Manuel and Sherson, Jacob F and Cheneau, Marc and Schau{\ss}, Peter and Fukuhara, Takeshi and Bloch, Immanuel and Kuhr, Stefan},
    number = {7338},
    month = {3},
    pages = {319--324},
    volume = {471},
    publisher = {Nature Publishing Group},
    url = {https://doi.org/10.1038/nature09827},
    isbn = {1476-4687 (Electronic){\textbackslash}r0028-0836 (Linking)},
    doi = {10.1038/nature09827},
    issn = {0028-0836},
    pmid = {21412333},
    arxivId = {1101.2076}
}

\newpage
\appendix
\section{Optimization results of post-imaging RSC}
\label{Parameter_table_of_post-imaging_RSC}
The parameters optimized for post-imaging Raman sideband cooling~(RSC) using Bayesian optimization are summarized in table~\ref{table:RSC_param}.
The beam configuration and notation follow those described in our previous work~\cite{Yamamoto2020Single-site-resolvedLattice}.

The single-photon detuning of the Raman beams is $\Delta/2\pi \simeq -100~\mathrm{GHz}$.
Under the optimized conditions, the effective two-photon Rabi frequencies for the carrier transitions are $\Omega_H/2\pi=25~\mathrm{kHz}$ and $\Omega_D/2\pi=24~\mathrm{kHz}$, respectively.

\begin{figure}[h]
    \centering
    \includegraphics[width=0.5\linewidth]{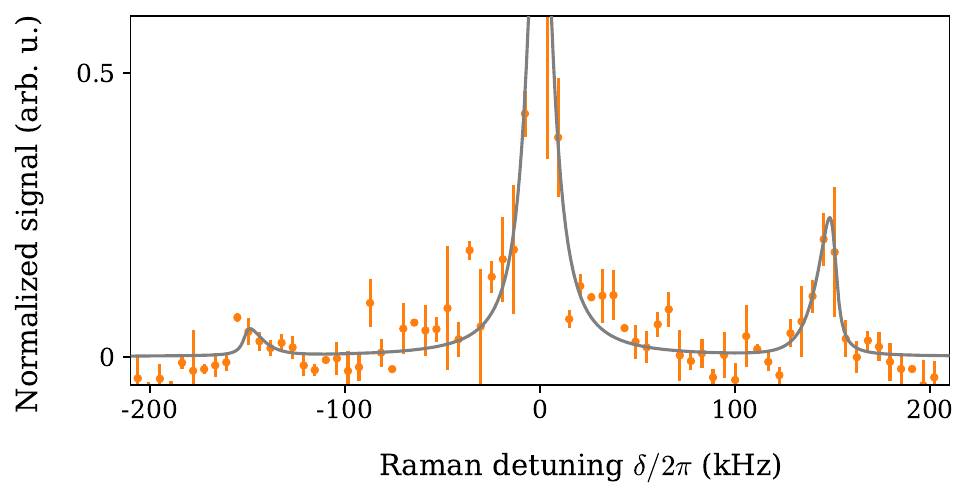}
    \caption{
    Raman sideband spectrum measured using the R1(H)-R2 Raman transition after post-imaging RSC with a cooling duration of $t_\mathrm{RSC}=150~\mathrm{ms}$.
    The data points show the measured normalized signal, while the solid curve represents a fit to the spectrum.
    }
    \label{fig:Raman_sideband_spectroscopy}
\end{figure}

To characterize the motional state of the atoms after post-imaging RSC, we performed Raman sideband spectroscopy using the R1(H)-R2 Raman transition.
Figure~\ref{fig:Raman_sideband_spectroscopy} shows the spectrum obtained after a cooling duration of $t_\mathrm{RSC}=150~\mathrm{ms}$.
From the sideband asymmetry, we estimate the mean motional quantum number to be $\langle n\rangle=0.2^{+0.3}_{-0.2}$, corresponding to a motional ground-state population of approximately $83\%$ assuming a thermal distribution.

\begin{table}[h]
    \centering
    \caption{
    Optimized parameters for post-imaging RSC.
    The intensities $I$ of the optical-pumping beams (OP1 and OP2) and Raman beams (R1(H), R1(D), and R2), together with two-photon detunings $\delta$ of the Raman transitions, were optimized using Bayesian optimization.
    The detunings listed for R1(H) and R1(D) represent the two-photon detunings of the respective Raman transition.}
    \begin{tabular}{@{}ccc}
    \br
    &   $I$~($\mathrm{W/m^2}$)& $\delta/2\pi$~(kHz)\\
    \hline
    OP1&    $4.71$&         --\\
    OP2&    $3.06$&         --\\
    R1(H)&  $731$&         $-123.8$\\
    R1(D)&  $659$&         $-55.3$\\
    R2&     $185 \times 10^3$&         --\\
    \br
    \end{tabular}
    \label{table:RSC_param}
\end{table}

\section{Lattice-site displacement due to relative phase drift}
\label{Lattice_Phase_shift}

\begin{figure}[h]
    \centering
    \includegraphics[width=0.5\linewidth]{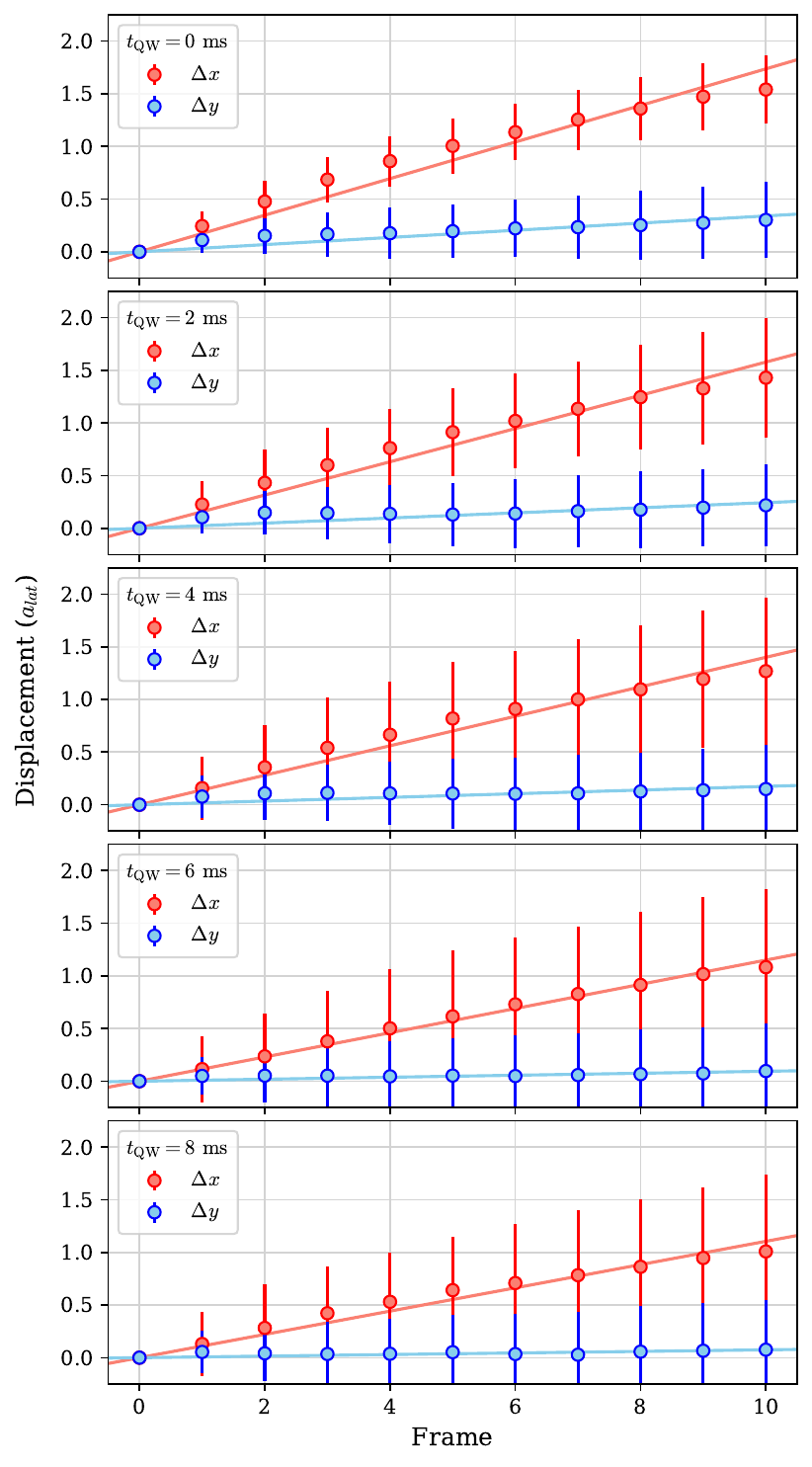}
    \caption{
    Lattice-site displacement due to relative phase drift between the lattice beams for different hold times during quantum-walk measurements (from 0~ms in the top panel to 8~ms in the bottom panel). The horizontal axis, labeled "Frame", indicates the frame number relative to the first image. 
    The maximum lattice-site drift rates are $0.174(6)~a_{\mathrm{lat}}/\mathrm{frame}$ and $0.034(2)~a_{\mathrm{lat}}/\mathrm{frame}$ along the x and y directions, respectively.
    The $x$ and $y$ axes correspond to the horizontal and vertical directions in the imaging plane, respectively.
    }
    \label{fig:site_drift}
\end{figure}

In our apparatus, the relative phases between the laser beams forming the triangular optical lattice drift over time, causing shifts in the lattice-site positions (figure~\ref{fig:site_drift}).
Such shifts can introduce systematic errors in the measured atomic displacements and consequently affect the performance of sequential measurements.
To account for this effect, we determine the lattice-site displacements in the second through eleventh images relative to those in the first image and correct the measured atomic positions accordingly.

The maximum evaluated site-drift rate is $0.174(6)~a_{\mathrm{lat}}/\mathrm{frame}$.
No significant dependence of the drift rate on the hold time for quantum dynamics was observed.

\section{Tunneling energies in an imbalanced triangular lattice}
\label{Calc_tunneling_energies}

\begin{figure}[h]
    \centering
    \includegraphics[width=1\linewidth]{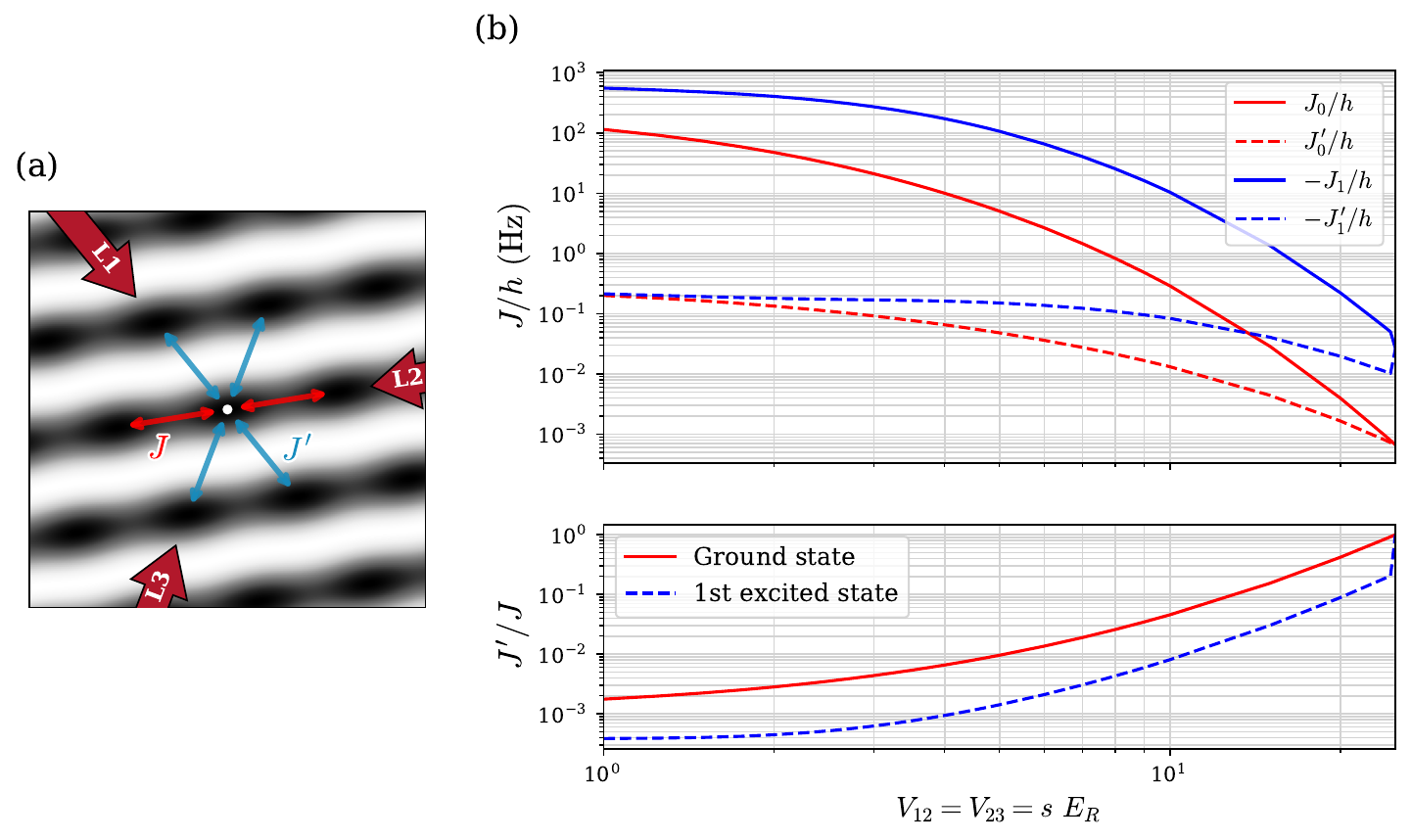}
    \caption{
    (a) Potential landscape of an effectively one-dimensional optical lattice with $V_{12}=V_{23} = 2.5~E_R$ and $V_{31} = 25~E_R$.
    The large dark-red arrows labeled $L_1$, $L_2$, and $L_3$ indicate the propagation directions of the three lattice beams, while the small bright-red and blue arrows indicate the tunneling directions along and transverse to the one-dimensional lattice, respectively.
    (b) Calculated tunneling energies as a function of $V_{12}=V_{23}$, with $V_{31}$ fixed at $25~E_R$. 
    The red and blue curves correspond to atoms in the motional ground and first excited states, respectively.
    Solid and dashed lines represent tunneling along the one-dimensional and transverse directions, respectively.
    For the first excited state, $-J_1/h$ and $-J'_1/h$ are plotted because the tunneling energies are negative.
    The lower panel shows the ratio of transverse to longitudinal tunneling energies, $J'/J$, for the motional ground state (red solid curve) and first excited state (blue dashed curve).
    }
    \label{fig:imbalanced_tunneling_rate}
\end{figure}

A triangular optical lattice formed by three laser beams can be characterized by the potential depths associated with the three beam pairs, $V_{12}$, $V_{23}$, and $V_{31}$. 
An array of one-dimensional lattice systems can be realized when two of the three beam-pair potentials are much shallower than the third~(figure~\ref{fig:imbalanced_tunneling_rate}(a)).
We therefore choose $V_{12}=V_{23}\ll V_{31}$, such that the resulting one-dimensional direction minimizes the influence of the external harmonic confinement.
For this configuration, we calculate the tunneling energies as a function of $V_{12}~=V_{23}$ while keeping $V_{31}=25~E_R$ fixed, as shown in figure~\ref{fig:imbalanced_tunneling_rate}(b).
For the experimental parameters $V_{12}=V_{23}=2.5~E_R$ and $V_{31}=25~E_R$, the calculated nearest-neighbor tunneling energies for atoms in the motional ground state are $J_{0}/h=31.3~\mathrm{Hz}$ along the one-dimensional direction and $J_{0}^\prime/h=0.11~\mathrm{Hz}$ in the transverse direction.
The corresponding tunneling energies for atoms in the first excited motional states are $J_{1}/h=-335~\mathrm{Hz}$ and $J_{1}^\prime/h=-0.17~\mathrm{Hz}$, respectively.

\end{document}